\documentstyle[prd,preprint,tighten,aps]{revtex}
\begin {document}
\draft
\title{
On the balance of the solar p--p chain}
\author{Attila Cs\'ot\'o\cite{email}}
\address{National Superconducting Cyclotron Laboratory,
Michigan State University, East Lansing, Michigan 48824}
\date{May 24, 1995}

\maketitle

\begin{abstract}
\noindent
I show that the solar neutrino fluxes, predicted by standard
nuclear and solar physics, can get closer to the experimental
observations if one takes the freedom to introduce two free parameters
into the model, as it is done in the MSW solution. I point out
that the plasma electron capture of $^7$Be and the
$^3$He($^3$He,$2p$)$^4$He nuclear reactions deserve further
experimental and theoretical attention.
\end{abstract}

\narrowtext

All four currently operating solar neutrino experiments, Homestake,
Kamiokande, SAGE, and GALLEX, show a deficit of the solar neutrinos
reaching the earth \cite{exp}. As these detectors have different
energy thresholds, they are sensitive to different parts of the
neutrino spectra. Table I shows the observed neutrino capture rates of the
Homestake and gallium detectors compared with the predictions of
Bahcall \cite{Bahcall}. The ratio of the measured Kamiokande $^8$B
neutrino flux to the prediction of Ref.\ \cite{Bahcall} is
0.51$\pm$0.04$\pm$0.06. We can see that the degree of deficit is
rather different in the different type of experiments. Model
independent analyses, which use only the luminosity constraint,
have shown the following \cite{modind}.

(i) The results of the three different types of experiments are hardly
compatible with each other within the standard nuclear-, solar-, and
neutrino model. If we take the $^7$Be ($\phi_7$), and $^8$B ($\phi_8$)
neutrino fluxes as free parameters, then the best (but still not very
good) fit to the three data would give a 50\% reduction
in the $\phi_8$ flux, and a practically zero $\phi_7$ flux \cite{fit}.
The major contradiction seems to be between the Homestake and Kamiokande
results. The deficit is much more severe in the former experiment, despite
the fact that this experiment has contributions from neutrinos other
than $^8$B.

(ii) The $^8$B neutrinos alone cannot be responsible for the solar
neutrino deficit. This is because the gallium experiments are hardly
sensitive to $\phi_8$. It turns out that to get closer to the
experiments, both $\phi_7$ and $\phi_8$ should be suppressed, but this
suppression is much stronger in $\phi_7$. It excludes the possibility
that the inaccurate $^7$Be($p,\gamma$)$^8$B cross section is the
major source of the problem. In fact, the recently suggested low values
of this cross section \cite{be7p} would increase the theoretical
$\phi_7/\phi_8$ ratio, thus even exaggregating the solar neutrino
problem.

Currently, the favorite explanation of the solar neutrino problem is the
Mikheyev--Smirnov--Wolfenstein (MSW) effect \cite{MSW}. It assumes that
the weakly interacting neutrino eigenstates ($\nu_e$, $\nu_\mu$,
$\nu_\tau$), which are linear combinations of the mass eigenstates,
can transform into each other while interacting with the
solar matter. In the two-component oscillation model the $\nu_e$
neutrinos are converted into $\nu_\mu$ or $\nu_\tau$ with the mixing
angle $\theta$ and mass difference $\Delta m$. It can be shown
that for certain values of $\theta$ and $\Delta m$ the theoretical
neutrino flux predictions of Table I can be reduced to the experimental
values in all experiments, simultaneously, see, e.g.,\ Ref.\ \cite{Gal}.

I would like to emphasize, however, that although the MSW mechanism
is an exciting possibility, if we forget about its mathematical and
particle physics pedigree, it is nothing but introducing two free
parameters to fit three data. Moreover, these free parameters are
introduced in a clever way, as they do not have any feedback on the
energy generation of the Sun, thus $\theta$ and $\Delta m$ can be
chosen without any constraint. Also, there are reserves in this
mechanism. If one wants to describe further neutrino experiments
(e.g.\ atmospheric-, reactor-, etc. experiments) and the above scheme
does not work, one still can introduce the third neutrino flavor, with
the additional mixing angles and mass differences as free parameters.
However, as the number of independent data is larger and larger, we are
running out of free parameters, and have to introduce other exotics,
such as sterile neutrinos, inverted mass hierarchy, etc.\ \cite{steril}.

If we allow ourselves to introduce two free parameters into the
model, we can do it in a simpler way, without introducing a new
mechanism, beyond the Standard Model. We can keep the standard
nuclear-, solar-, and neutrino physics, and introduce these parameters
into such nuclear reactions, whose cross sections have never been
measured in the interesting energy range. I emphasize, that the
following is currently only a pure theoretical possibility, without any
supporting experimental evidence or theoretical reaction model. But
currently the same is true for the MSW mechanism.

A careful survey of the $p-p$ chain suggests, that the best places
to introduce free parameters are the $^7$Be($e^-,\nu$)$^7$Li
electron capture from the solar plasma, and the $^3$He($^3$He,$2p$)$^4$He
reaction. The capture rate of the $^7$Be($e^-,\nu$)$^7$Li reaction
is thought to be known with high precision because its knowledge
requires only ``weak interaction theory and the local physical condition
of the solar plasma'' \cite{Bahcall}. Let me point out, however, that
there are cases where only standard atomic physics takes places, and
still there are factor of two differences between theory and experiment.
Such a case is, for example, the electron screening in low-energy
nuclear reactions \cite{screening}. Without any experiment, we cannot
be {\it a priori} sure that the $^7$Be($e^-,\nu$)$^7$Li capture rate
is correct. If this rate were smaller than the current
theoretical value by a factor of two, then it would decrease the $\phi_7$
flux and increase the $\phi_8$ flux by the same amount.

It was suggested twenty years ago, that a low-energy resonance in the
$^3$He($^3$He,$2p$)$^4$He reaction would suppress both the $\phi_7$
and $\phi_8$ neutrino fluxes \cite{Fowler}. The latest experiment
went down to 25 keV energy without observing a resonance \cite{Krauss}.
The LUNA experiment, currently running at Gran Sasso Laboratory, is
planning to reach 15 keV \cite{Luna}. At such low energies, deep
below the Coulomb barrier, the properties of such a hypothetical
resonance are solely determined by the Coulomb penetration. It can be seen,
that such a resonance, if existed somewhere betwen 15 keV and {\it zero}
energy, would cause the suppression of the $\phi_7$ and $\phi_8$
fluxes by a factor of 4.3 and 3.3, respectively \cite{Castellani}. (This
is in the case of a $0^+$ resonance; other resonant partial waves could
cause even higher suppression.) It means, that even if the experiments
could reach 10 keV by heroic efforts, the existence of a state, which
would have strong influence on the neutrino fluxes, is still possible
below this energy.

The combined effect of the above-mentioned two changes in the cross
sections would be a reduction of $\phi_7$ by a factor of 8.6 and
that of $\phi_8$ by a factor of 1.6. Although the agreement with
the solar neutrino experiments would not be perfect, all theoretical
predictions would move toward the right direction, and the
$\phi_7/\phi_8$ ratio would considerably decrease. I note, that these
changes of the neutrino fluxes are only first order estimates,
because they neglect the feedback to other reactions of the $p-p$
chain. The correct way to calculate the effects of the changes of
the nuclear cross sections would be to use these cross sections in
a solar model. Such a study is in progress.

In conclusion, I have emphasized, that from practical point of
view, currently the MSW solution of the solar neutrino problem is
nothing, but introducing two free parameters to fit three data.
If we allow ourselves this amount of freedom within the nuclear physics
part of the problem, we could also get considerably closer to the
experimental results, without going beyond the Standard Model. The most
interesting reactions of the solar $p-p$ chain, from this point of view,
are the $^7$Be($e^-,\nu$)$^7$Li electron capture from the solar plasma,
and the $^3$He($^3$He,$2p$)$^4$He process. These reactions should deserve
further theoretical and exprimental studies. My preliminary theoretical
studies of the $^3$He($^3$He,$2p$)$^4$He reaction show, for example,
that the dynamics of this process is much more complex than the existing
models have assumed so far, and involves dynamical degrees of freedom of
the six-nucleon problem, that were not included before \cite{Csoto}.

\mbox{}

This work was supported by NSF Grant Nos. PHY92-53505 and
PHY94-03666.

\begin{table}
\caption{Neutrino capture rates in the Homestake and gallium
experiments.}
\begin{tabular}{cr@{}lr@{}l}
 & \multicolumn{4}{c}{Capture rate (SNU)} \\ \cline{2-5}
Neutrino source & \multicolumn{2}{c}{Homestake} & \multicolumn{2}{c}{GALLEX
\& SAGE} \\
\tableline
pp        & 0.&0   & 70.&8   \\
pep       & 0.&2   &  3.&0   \\
hep       & 0.&03  &  0.&06  \\
$^7$Be    & 1.&1   & 34.&3   \\
$^8$B     & 6.&1   & 14.&0   \\
$^{13}$N  & 0.&1   &  3.&8   \\
$^{15}$O  & 0.&3   &  6.&1   \\
$^{17}$F  & 0.&003 &  0.&06  \\
\tableline
Total    & 7.9$\pm$&0.87 & \multicolumn{2}{c}
 {\ \ \ $132^{+6.7}_{-5.7}$} \\
Experiment & 2.55$\pm$&0.25 &  79$\pm$10$\pm$6 \hspace{1mm} &
  \hspace{1mm} $74^{+13+5}_{-12-7}$
\end{tabular}
\end{table}

\end{document}